\documentstyle[twoside,StarsAndGalaxies,epsf]{article}
\newcommand{\angst}{\buildrel _{\circ} \over {\mathrm{A}}}

\begin{document}
\setcounter{volume}{1}
\setcounter{spage}{1}
\setcounter{epage}{17}
\setcounter{year}{2018}
\thispagestyle{headings}
\setcounter{page}{\value{spage}}
\title{Radial Velocity Survey of Nearby OB Stars}
\author{Yuki \textsc{Moritani}}
\affiliation{The University of Tokyo, Kavli Institute for the Physics and Mathematics of the Universe (WPI), \\
The University of Tokyo Institutes for Advanced Study, \\
5-1-5, Kashiwa, Chiba 277-8583, Japan}
\email{yuki.moritani@ipmu.jp}
\moreauthors{Takuma \textsc{Suda}, Toshikazu \textsc{Shigeyama}}
\affiliation{The University of Tokyo, Research Center for the Early Universe, \\
Hongo 7-3-1, Bunkyo, Tokyo 113-0033, Japan}
\moreauthors{Takayuki R. {\sc Saitoh}}
\affiliation{Tokyo Institute of Technology, Earth-Life Science Institute, \\ 
2-12-1, Ookayama, Meguro, Tokyo, 152-8551, Japan}
\radate{2018 October 31}{2018 December 4}

\begin{abstract}
We report on the current status of the radial velocity monitoring of nearby OB stars to look for binaries with small mass ratios.
The combined data of radial velocities using the domestic 1-2 m-class teleoscopes seems to confirm the variations of radial velocities in a few weeks for four out of ten target single-lined spectroscopic binaries.
More data are needed to estimate the exact periods and mass distributions.
\end{abstract}


\section{Introduction}
Binary stars are important probes for the understanding of stellar evolution.
Extensive surveys on the characteristics of binary stars have revealed the period distribution and the mass ratio distribution of binaries \cite{Duquennoy1991,Latham2002,Raghavan2010}, which provides useful information on star formation and chemical evolution.
Thanks to these large-scale surveys, the number of samples have increased to more than 1,000, while the survey sample is, in most cases, restricted to binaries whose primaries are solar-type stars or low-mass stars at the main sequence phase.
For massive stars, on the other hand, the number of stars is limited, but it is suggested that the census of massive star binaries are quite different from low-mass stars.
Most of massive binaries are thought to undergo interactions between two stars via roche-lobe overflow, envelope stripping, or merging during their evolution \cite{Sana2012}.

The identification of massive star binaries have some limitations in observations.
Surveys of binaries using radial velocity monitorings, proper motion, or other methods rely on existing bright or nearby stars.
It is difficult to determine the nature of binary systems for faint or distant stars that are located in the Galactic halo and should have low metallicity.
The mulitplicity fraction at lower metallicity suffers large uncertainties with the current surveys \cite{Rastegaev2010,Moe2018}.
Moreover, it is impossible to determine directly the properties of massive star binaries that have already ended their lives. 
Therefore, the survey of binary systems for massive (O and early B type) stars can be made only for metallicity comparable to the solar value and for stars in the solar neighborhood.

In this study, we report on the survey of OB star binaries based on the monitoring of their radial velocities.
In particular, we focus on the binaries with small mass ratios, $q = M_2 / M_1$ where $M_1$ and $M_2$ is the mass of primary and secondary, respectively, to compensate for the scarcity of the data.
One of the examples of the confirmed OB star binaries with small mass ratios is HD 164438 \cite{Mayer2017} whose mass ratio and period is 0.1 -- 0.22 and 10.25 days, respectively.
The details of the observations are given in \S~\ref{sec:obs}.
The current results are reported in \S~\ref{sec:res}, followed by brief summary in \S~\ref{sec:sum}.

\section{Observations}\label{sec:obs}
Our target stars should have binary companions with the mass of $\sim$ 1 \msun\ since the primary stars are expected to have $\sim$ 10 \msun.
If the orbital periods of our target stars are in the range from a few days to a few years, the amplitude of the radial velocities should be 1 -- 100 ${\rm km \; s}^{-1}$.
This periodic variation is detectable by the monitoring of the radial velocities using the observational setups as described below.

\subsection{Target Selection}
The program stars have been selected from 64112 OB stars listed in the Catalogue of Stellar Spectral Classifications \cite{Skiff2014}.
Among them, we have excluded extra Galactic sources and binaries already identified to be double-lined spectroscopic binaries (SB2) [e.g., the 9th Catalogue of Spectroscopic Binary Orbits \cite{Pourbaix2004}], ecllipse binaries \cite[for instance]{Malkov2006}, astrometry binaries [e.g., the Washington Double star Catalogue \cite{Mason2001}, or visual binaries (the Sixth Catalog of Orbits of Visual Binary Stars\footnote{https://www.usno.navy.mil/USNO/astrometry/optical-IR-prod/wds/orb6}), and High Mass X-ray Binaries \cite{Liu2006}.
We have checked more than 30 papers and catalogues to search for identified binaries.
Besides, we have also explored literatures to update binary parameters as much as posible.

Among the rest of 56370 stars in the catalogues that potentially include single-lined spectroscopic binaries (SB1), we have selected 59 SB1 candidates as target stars.
This is because the optical spectra should be dominated by the OB stars that are much brighter than the late-type companion stars.
As a first step of our survey, we have selected ten stars identified as SB1 by \cite{Chini2012}, which can be monitored feasibily using the domestic 1--2 m-class telescopes.
These ten stars are sufficiently bright (V $\leq$ 8) and are located in the northern sky ($\delta > -25^\circ $).
Our tentative target stars are given in Table~\ref{tab:target}.

It is to be noted that HD 164438 was initially included in our sample, but during the course of our observations, the orbital information on this star was determined \cite{Mayer2017}.
We found that their results are qualitatively consistent with those of our tentative analysis, and decided to drop the star from the target list.

\begin{table}[h]
\caption{Target Stars}
\begin{center}
\begin{tabular}{lllll}
		\hline
		Target Name	& RA	& Dec	& V Mag.	& Sp. Type \\ \hline \hline
		HD 18326	& 02 59 23.17	& +60 33 59.5	& 8.0	& B0 \\
		HD 35439	& 05 24 44.83	& +01 50 47.2	& 4.9	& B3e \\
		HD 52266	& 07 00 21.08	& -05 49 36.0	& 7.2	& OB \\
		HD 93521	& 10 48 23.51	& +37 34 13.1	& 7.1	& B3$^a$ \\
		HD 98664	& 11 21 08.19	& +06 01 45.6	& 4.1	& B9 \\
		HD 184279	& 19 33 36.92	& +03 45 40.8	& 7.0	& B2e \\
		HD 190864	& 20 05 39.80	& +35 36 28.0	& 7.8	& O6 \\
		HD 194739	& 20 27 02.14	& +09 05 31.7	& 7.5	& B2.5V \\
		HD 195592	& 20 30 34.97	& +44 18 54.9	& 7.1	& B1e \\
		HD 204827	& 21 28 57.76	& +58 44 23.2	& 8.0	& O9.5IV \\
		\hline
\end{tabular}\label{tab:target}
\end{center}
Note: (a) \cite{Rauw2008} reported the spectral type is O9.5V.
\end{table}

\subsection{Instruments and Dara Reduction}\label{sec:ins}
We have conducted a monitoring of the ten program stars since 2016 using the domestic 1--2m-calss telescopes.
We used a medium-resolution spectrograph to detect shorter-period ($\mathrm{P_{orb}}$ of day--week) binaries, and high-resolution spectrogrphs to detect longer-period ($\mathrm{P_{orb}}$ of years) binarirs with the estimated variations of the radial velocities of 1--100 $\mathrm{km \;s^{-1}}$.

Medium-resolution spectra were obtailed with Medium And Low-dispersion Long-slit Spectrograph (MALLS) \cite{Ozaki2005} equipped to the Nayuta 2.0\,m telescope at the Nishi Harima Astronomical Observatory (NHAO).
To cover 4400--5000 $\angst$, where many metal lines exist in addition to strong HI and HeI lines, we have set two center wavelengths of the 1800\,mm$^{-1}$ grating; 4400 $\angst$ and 4800 $\angst$. 
The spectral resolution is $R \sim$ 5000.
Most of the spectra produce $S/N$ higher than 200.

High-resolution spectra were obtained using two spectrographs; High Dispersion Echelle Spectrograph (HIDES) \cite{Izumiura1999,Kambe2013} equipped to the 188 cm telescope at the Okayama Astrophysical Observatory (OAO), and Gunma Astronomical Observatory Echelle Spectrograph (GAOES) \cite{Hashimoto2006} equipped to the 1.5 m telescope at the Gunma Astronomical Observatory (GAO).
Both spectra cover wavelength range of 4000--5000 $\angst$ with a single grating configuration.
The spectral resolution of HIDES and GAOES is $R\sim 60000$ and $R\sim 30000$, respectively.
Most of the spectra produce $S/N$ higher than 200.

All the spectra were reduced using the IRAF\footnote{http://iraf.noao.edu/} in the standard way --- bias subtraction, flat fielding, extraction to 1-d spectrum, wavelength calibration, intensity normalization, and heliocentric correction.

Radial velocity variations have been estimated by the cross-correlation method.
We have used the IRAF {\tt rv} package to measure the shift of the radial velocity from the reference spectrum.
Here, the reference spectum was chosen from one of the high-resolution spectra by OAO/HIDES and GAO/GAOES.
The {\tt FXCOR} method of {\tt rv} package calculates the cross-correlation value of the two spectra by shifting one spectra in pixels. The peak of pixel shift corresponds to our derived relative velocity.
A Gaussian fit was made to the center of the peak in individual spectrum.
The error is estimated by the width of half the power of the fitted peak, i.e., the FWHM of the fitting.

\section{Results}\label{sec:res}
\begin{figure}[htbp]
 \epsfxsize=12.0cm
 \centerline{\epsffile{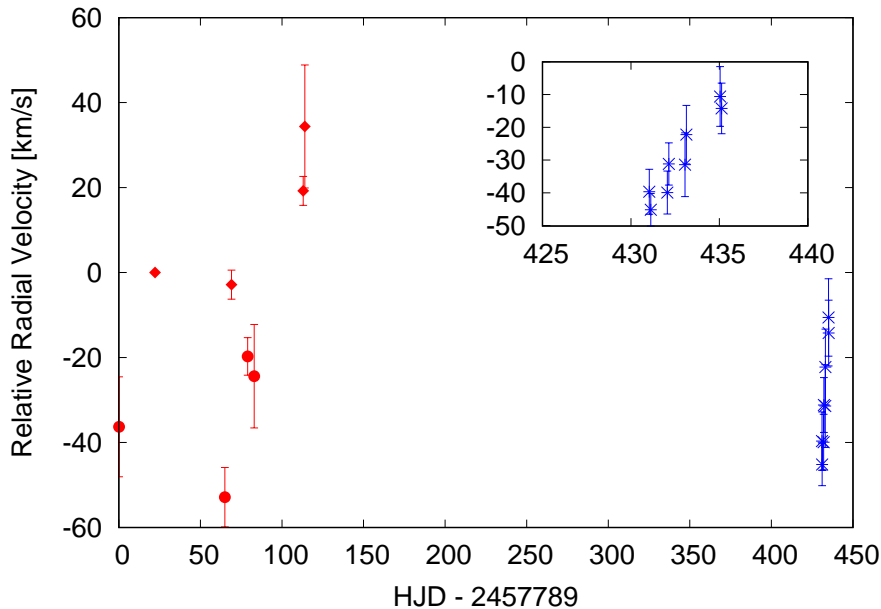}}
\caption{Monitoring history of the radial velocity of HD 93521. 
The zero point of the day is adjusted to the second observation of our program because the first observational datum has a large uncertainties caused by low signal to noise ratio. The radial velocity on the fiducial observation is also set at zero without error bars because it is chosen as the reference value in the cross-correlation method. The data with high-resolution spectroscopy using OAO or GAOES are shown in red circles, while those with medium-resolution spectroscopy using MALLS are shown in blue crosses. The box in the top-right corner is the magnification for the data obtained in short duration.}\label{fig:93521}
\end{figure}

\begin{figure}[htbp]
 \epsfxsize=12.0cm
 \centerline{\epsffile{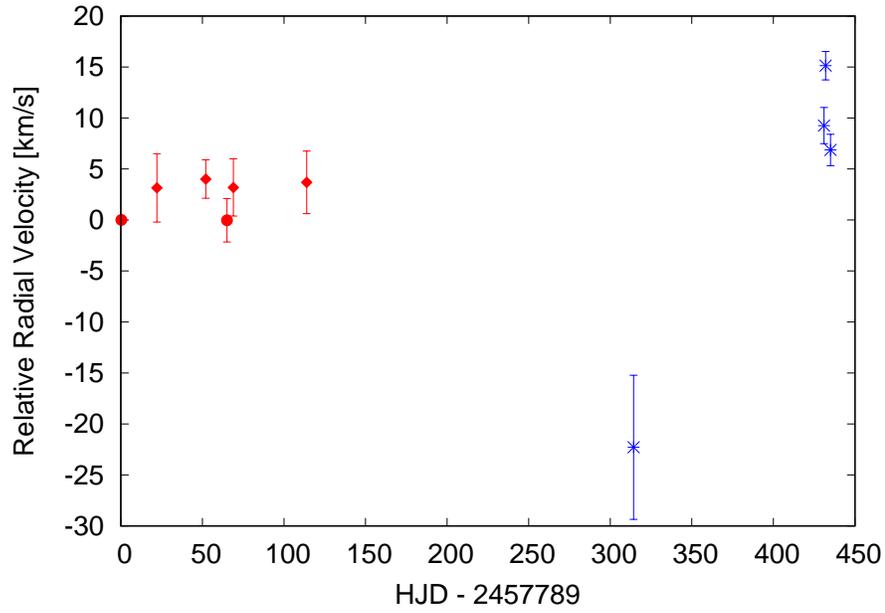}}
\caption{The same as in Fig.~\ref{fig:93521}, but for HD 98664. The zero point of the day is adjusted to the initial observation of our program.}\label{fig:98664}
\end{figure}

\begin{figure}[htbp]
 \epsfxsize=12.0cm
 \centerline{\epsffile{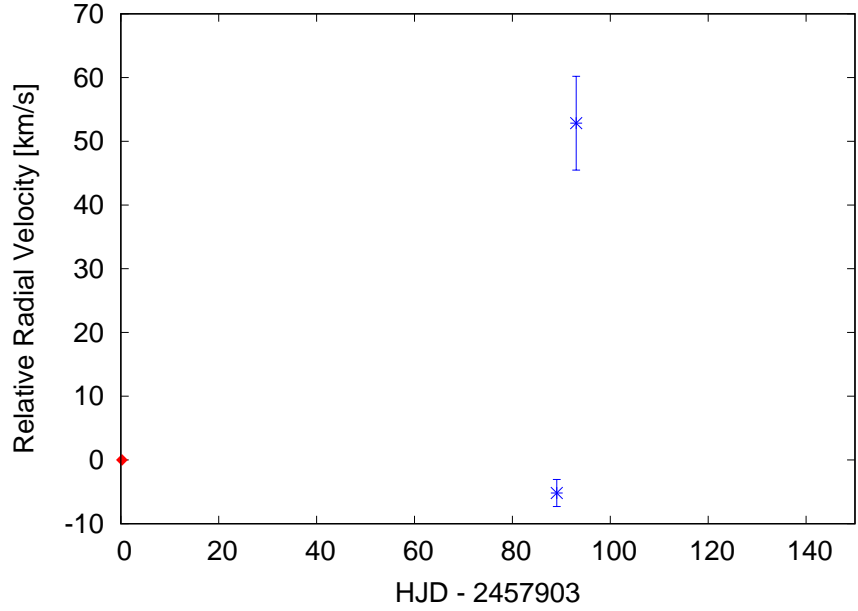}}
\caption{The same as in Fig.~\ref{fig:98664}, but for HD 184279.}\label{fig:184279}
\end{figure}

\begin{figure}[htbp]
 \epsfxsize=12.0cm
 \centerline{\epsffile{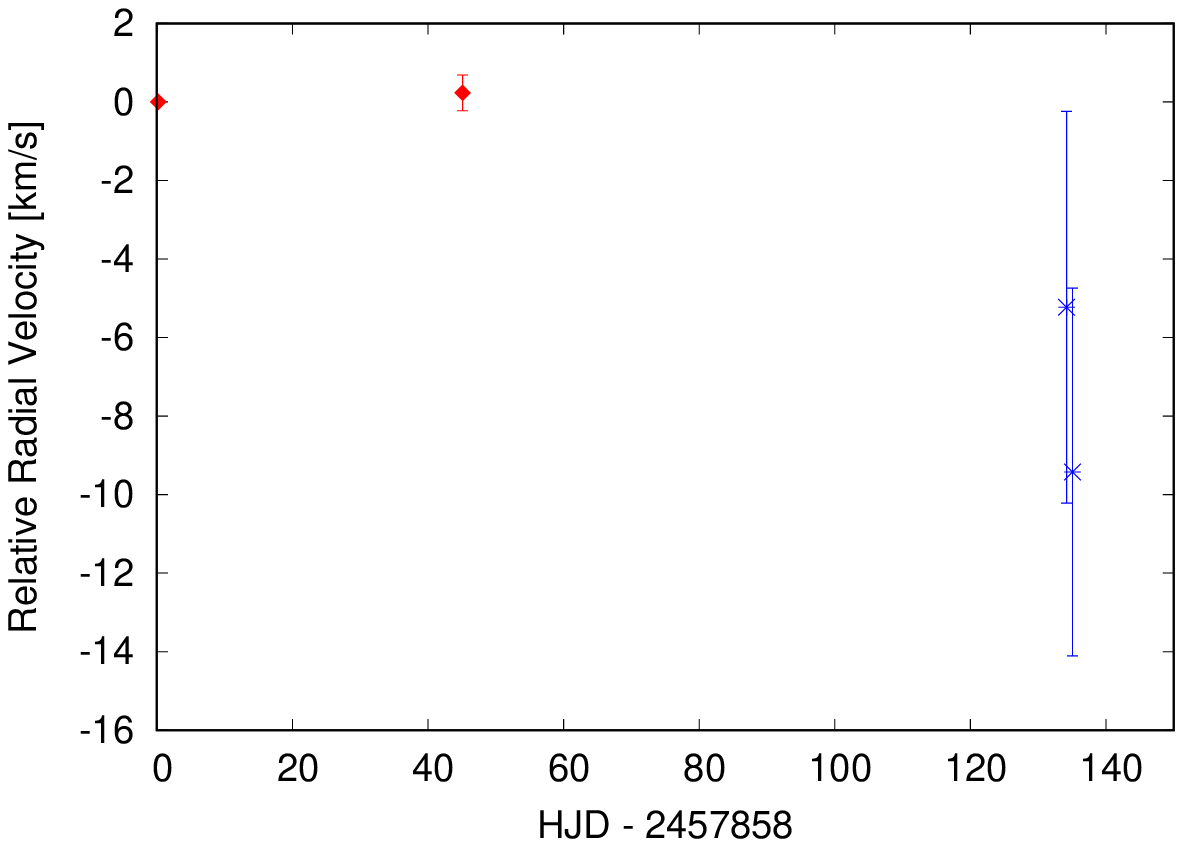}}
\caption{The same as in Fig.~\ref{fig:98664}, but for HD 190864.}\label{fig:190864}
\end{figure}

Figures \ref{fig:93521} -- \ref{fig:190864} show the variations of radial velocities during the period of the observations for HD 93521, HD 98664, HD 184279, and HD 190864, respectively.
These selected targets apparently show some variations in the radial velocities.
This is almost consistent with the expected rate of binaries using the Monte Carlo simulations of binary population \cite{Suda2013}.
For the simulations of binary population synthesis, the initial mass function is assumed to be a Salpeter's power law function.
The period distribution is based on the compilation of the literature on OB star binaries as described in the previous section.
Binary population synthesis predicts that one out of six SB1 stars have mass ratios of the order of 0.1, which corresponds to the binary period of a few days to a few weeks.

HD 93521 (Figure \ref{fig:93521}) is known to be a peculiar star among our target stars, and needs some attention in the current survey.
The star is an O9.5V type star, located at high-latitude in the Galaxy.
A very rapid rotation ($\sim$ 390 $\mathrm{km \; s^{-1}}$) was reported for HD 93521 \cite{Rauw2008}.
It is confirmed that there are 1.75 hr and 2.89 hr periods caused by non-radial pulsations, by searching for periodicity using H$\alpha$ and He I lines.
In this study, the MALLS obserations imply that the radial velocity increases monotonously by $\sim$ 50 $\mathrm{km \; s^{-1}}$ in 5 days.
If this variation is caused by the biniariy, the orbital period is expected to be $\sim$ 1 month, assuming that the amplitude of the variability is $\pm 60~\mathrm{km \; s^{-1}}$ as shown in Fig.~\ref{fig:93521}.

No observational constraints on biniary parameters have been reported for the other three candidates shown in the figures.
To confirm the binarities and determine the binary periods, we need radial velocity data in appropriate intervals.

\subsection{Uncertainties}
The errors of individual measurement in Figures \ref{fig:93521} -- \ref{fig:190864} are estimated by {\tt FXCOR} method as shown in \S~\ref{sec:ins}.
The uncertainties associated with the fitting procedure are much smaller than the variations in the measured radial velocity for HD 93521 as long as we use the same spectrograph.
For HD 98664 and HD 184279, the results with MALLS clearly show the large variation well beyond the uncertainty of the estimated radial velocities.

Systematic differences among different facilities are also investigated by the comparisons of the line profiles of the same star in the same day.
The line profiles obtained by different spectrographs show that the wavelength calibration is consistent with each other within the wavelength resolution of individual instruments.
The systematic offsets in radial velocity using different instruments should be less than several $\mathrm{km \; s^{-1}}$.
Therefore, we can safely conclude that the variations of the radial velocities are real even for HD 190864.
It is to be noted that this kind of comparisons is also made between HIDES and GAOES, for instance, by the monitoring of the line profile variability in the Balmer lines of V725 Tau \cite{Moritani2013}.
The same monitoring has been performed for the same target using the MALLS data (Okazaki et al., in prep.).

\section{Summary}\label{sec:sum}
The paper reports on the preliminary results for the monitoring of radial velocities for ten OB stars that are known to be single-lined spectroscopic binaries.
The purpose of this study is to identify the OB star binaries with small mass ratios and to determine their periods and mass ratios.
We have obtained spectra from Medium And Low-dispersion Long-slit Spectrograph (MALLS) on the Nayuta 2.0\,m telescope at the Nishi Harima Astronomical Observatory, High Dispersion Echelle Spectrograph (HIDES) on the 188 cm telescope at the Okayama Astrophysical Observatory, and Gunma Astronomical Observatory Echelle Spectrograph (GAOES) on the 1.5 m telescope at the Gunma Astronomical Observatory (GAO).
The cross correlation method to confirm the variations of the radial velocities allow us to conclude that at least four of our target stars show evidence of short period binaries with a few days to a few weeks.
Further monitoring of the target stars will constrain the binary parameters.

\acknowledgments

This work has been supported by a Grant-in-Aid for Scientific Research (JP16K05287), from the Japan Society of the Promotion of Science.


\begin{thebibliography}{99}

\bibitem[Duquennoy et al. (1991)]{Duquennoy1991}
Duquennoy, A., Mayer, M., \AA, 248, 485

\bibitem[Latham et al. (2002)]{Latham2002}
{Latham}, D.~W., {Stefanik}, R.~P., {Torres}, G., {Davis}, R.~J., {Mazeh}, T., {Carney}, B.~W., {Laird}, J.~B., {Morse}, J.~A., 2002, \AJ, 124, 1144

\bibitem[Raghavan et al. (2010)]{Raghavan2010}
{Raghavan}, D., {McAlister}, H.~A., {Henry}, T.~J., {Latham}, D.~W., {Marcy}, G.~W., {Mason}, B.~D., {Gies}, D.~R., {White}, R.~J., {ten Brummelaar}, T.~A., 2010, \ApJS, 190, 1

\bibitem[Sana et al. (2012)]{Sana2012}
Sana, H., de Mink, S. E., de Koter, A., Langer, N., Evans, C. J., Gieles, M., Gosset, E., Izzard, R. G., Le Bouquin, J.-B., Schneider, F. R. N. 2012, \Science, 337, 444

\bibitem[Rastegaev (2010)]{Rastegaev2010}
Rastegaev, D.~A., 2010, \AJ, 140, 2013

\bibitem[Moe et al. (2018)]{Moe2018}
Moe, M., Kratter, K. M., Badenses, C., arXiv:1808.02116

\bibitem[Mayer et al. (2017)]{Mayer2017}
{Mayer}, P., {Harmanec}, P., {Chini}, R., {Nasseri}, A., {Nemravov{\'a}}, J.~A., {Drechsel}, H., {Catalan-Hurtado}, R., {Barlow}, B.~N., {Fr{\'e}mat}, Y., {Kotkov{\'a}}, L., 2017, \AA, 600, A33

\bibitem[Skiff (2014)]{Skiff2014}
Skiff, B.~A., 2014, VizieR Online Data Catalog,
http://vizier.u-strasbg.fr/viz-bin/VizieR?-source=B/mk
%
\bibitem[Pourbaix et al. (2004)]{Pourbaix2004}
Pourbaix, D., Tokovinin A.~A., Batten, A.~H. et al. 2004, \AA, 424, 727, \\
http://sb9.astro.ulb.ac.be/mainform.cgi
%
\bibitem[Malkov et al. (2006)]{Malkov2006}
Malkov, O.~Y., Oblak, E,. Snegireva, E.~A. and Torra, J. 2006, \AA, 446, 785
%
\bibitem[Mason et al. (2001)]{Mason2001}
Mason, B.~D., Wycoff, G.~L., Hartkopf, W.~I., Douglass, G.~G., and Worley, C.~E. 2001, \AJ, 122, 3466,
http://vizier.cfa.harvard.edu/viz-bin/VizieR-3?-source=B/wds,
%
\bibitem[Liu et al. (2006)]{Liu2006}
Liu, Q.~Z., van Paradijs, J. and van den Heuvel, E.~P.~J. 2006, \AA, 455, 1165
%
\bibitem[Chini et al. (2012)]{Chini2012}
Chini, R., Hoffmeister, V.~H., Nasseri, A., Stahl, O. and Zinnecker, H. 2012, \MN, 424, 1925
%
\bibitem[Ozaki \& Tokimasa (2005)]{Ozaki2005}
Ozaki, S. \& Tokimasa, N. 2005, Annual Report of the Nishi-Harima Astronomical Observatory (ISSN 0917-6926), No.~15, p.~15 - 29 (2005), 15, 15
%
\bibitem[Izumiura (1999)]{Izumiura1999}
Izumiura, H. 1999, Publications of the Yunnan Observatory, 77
%
\bibitem[Kambe et al. (2013)]{Kambe2013}
Kambe, E., Yoshida, M., Izumiura, H., et~al. 2013, \PASJ, 65
%
\bibitem[Hashimoto et al. (2006)]{Hashimoto2006}
Hashimoto, O., Malasan, H.~L., Taguchi, H., et~al. 2006, in The 9th Asian-Pacific Regional IAU Meeting, ed. W. Sutantyo, P.~W. Premadi, P. Mahasena, T. Hidayat, \& S. Mineshige, The 9th Asian-Pacific Regional IAU Meeting, 295
%
\bibitem[Suda et al. (2013)]{Suda2013}
{Suda}, T., {Komiya}, Y., {Yamada}, S., {Katsuta}, Y., {Aoki}, W., {Gil-Pons}, P., {Doherty}, C.~L., {Campbell}, S.~W., {Wood}, P.~R., {Fujimoto}, M.~Y., 2013, \MN, 432, L46
%
\bibitem[Rauw et al. (2008)]{Rauw2008}
Rauw, G., De Becker, M., van Winckel, H., Aerts, C., Eenens, P., Lefever, K., Vandenbussche, B., Linder, N., Naz{\'e}, Y. and Gosset, E. 2008, \AA, 487, 659
%
\bibitem[Moritani et al. (2013)]{Moritani2013}
Moritani, Y., Nogami, D., Okazaki, A. T., Imada, A., Kambe, E.; Honda, S., Hashimoto, O., Mizoguchi, S., Kanda, Y., Sadakane, K. and Ichikawa, K. 2013, \PASJ, 65, 83.
\end{thebibliography}
\end{document}